\documentclass[aps,prd,12pt,amssymb]{revtex4}

\begin{document}

\baselineskip=0.60cm

\newcommand{\ini}{\begin{equation}}
\newcommand{\fin}{\end{equation}}
\newcommand{\inir}{\begin{eqnarray}}
\newcommand{\finr}{\end{eqnarray}}
\newcommand{\inif}{\begin{figure}}
\newcommand{\finf}{\end{figure}}
\newcommand{\bc}{\begin{center}}
\newcommand{\ec}{\end{center}}

\def\ol{\overline}
\def\pa{\partial}
\def\ra{\rightarrow}
\def\ts{\times}
\def\df{\dotfill}
\def\bs{\backslash}
\def\dg{\dagger}

$~$

\hfill DSF-16/2003

\vspace{1 cm}

\title{INVERTING THE SEESAW FORMULA}

\author{D. Falcone}

\affiliation{Dipartimento di Scienze Fisiche,
Universit\`a di Napoli, Via Cintia, Napoli, Italy}

\begin{abstract}
\vspace{1cm}
\noindent
By inverting the seesaw formula we determine the heavy neutrino
mass matrix. The impact on the baryogenesis via leptogenesis and the
radiative lepton decays in supersymmetric models is described.
Links to neutrinoless double beta decay are also briefly discussed.
The analysis leads to two distinct matrix models. One has small mixings
while the other has one maximal mixing. Both cannot give a sufficient amount of
baryon asymmetry. Then we also comment on a different form of the Dirac
neutrino mass matrix, which does provide sufficient baryon asymmetry.
In a supersymmetric scenario the branching ratios of
radiative lepton decays are enhanced for this model.
\end{abstract}

\maketitle

\newpage

\section{Introduction}

The seesaw mechanism \cite{ss} is a simple framework which can explain
the smallness of neutrino mass.
It requires only a modest extension of the minimal standard
model, namely the inclusion of the heavy right-handed neutrino, but can be well
realized within left-right models \cite{lrm}, partial unified models \cite{pum},
and grand unified $SO(10)$ theories \cite{gut},
where the right-handed neutrino does exist.
Then the effective neutrino mass matrix $M_L$ is given by the seesaw formula
\ini
M_L \simeq M_{\nu} M_R^{-1} M_{\nu}^T,
\fin
where $M_R$ is the mass matrix of the right-handed neutrino and $M_{\nu}$ is the
Dirac neutrino mass matrix. The master formula (1) is valid when the
eigenvalues of $M_R$ are much larger than the elements of $M_{\nu}$ and in
such a case the eigenvalues of $M_L$ come out very small with respect to
those of $M_{\nu}$. Indeed, unlike $M_{\nu}$,
the generation of $M_R$ is not related to
the electroweak symmetry breaking and thus its scale may be very large.
Moreover, $M_R$ is a Majorana mass matrix, and as a consequence also $M_L$
is a Majorana mass matrix of left-handed neutrinos
(see for example \cite{bgg}). This fact is related to the violation of total
lepton number at high scale \cite{wein}, which should produce important phenomena
such as the baryogenesis via leptogenesis \cite{fy} and the
neutrinoless double beta decay \cite{furry}.
Lepton flavors are also violated, but in the nonsupersymmetric theory,
due to the smallness of neutrino mass, such processes are so suppressed to be
unobservable \cite{pet}, apart from neutrino oscillations. The situation is
different in the supersymmetric theory, even with universal soft breaking
terms, where some of these processes may be observable \cite{bm}.

Both lepton number and lepton flavor violations depend on the mass matrices
$M_{\nu}$ and $M_R$. On the other hand, we have several informations on the
effective neutrino mass matrix, coming from neutrino oscillations and more
generally from neutrino experiments. Therefore, it is reasonable, relating
$M_{\nu}$ to the charged fermions mass matrices, to obtain informations on
$M_R$ by inverting the seesaw formula,
\ini
M_R \simeq M_{\nu}^T M_L^{-1} M_{\nu}.
\fin
As a consequence, we should be able to determine also the impact on the
baryogenesis via leptogenesis, the neutrinoless double beta decay, and
for example the radiative lepton decays in some supersymmetric models.
The seesaw formula is valid above the $M_R$ scale, so that one should determine
$M_L$ at that scale. Although in several case the effect is not
relevant, we must take care of the renormalization issue (see the recent paper
\cite{frismi}).

In section II we discuss the Dirac mass matrices of quarks and leptons.
In section III we describe the effective neutrino mass matrix and in
particular its element $M_{ee}$, related to neutrinoless double beta decay.
In section IV we determine the mass matrix of right-handed neutrinos.
In section V and VI, respectively, we study the consequences for the baryogenesis
via leptogenesis and the radiative lepton decays in supersymmetry.
Finally, we present a discussion.

\section{Dirac mass matrices}

A symmetric form of the quark mass matrices, in agreement with the phenomenology
of quark masses and mixings, is described in Refs.\cite{cf,nmf}, and given by
\ini
M_d \simeq \left( \begin{array}{ccc}
0 & \sqrt{m_d m_s} & 0 \\
\sqrt{m_d m_s} & m_s & \sqrt{m_d m_b} \\
0 & \sqrt{m_d m_b} & m_b,
\end{array} \right)
\fin
\ini
M_u \simeq \left( \begin{array}{ccc}
0 & \sqrt{m_u m_c} & 0 \\
\sqrt{m_u m_c} & m_c & \sqrt{m_u m_t} \\
0 & \sqrt{m_u m_t} & m_t
\end{array} \right).
\fin
Moreover, in Ref.\cite{cf}, the charged lepton mass matrix has an analogous
form,
\ini
M_e \simeq \left( \begin{array}{ccc}
0 & \sqrt{m_e m_{\mu}} & 0 \\
\sqrt{m_e m_{\mu}} & m_{\mu} & \sqrt{m_e m_{\tau}} \\
0 & \sqrt{m_e m_{\tau}} & m_{\tau}
\end{array} \right).
\fin
Since the hierarchy and scale of charged lepton masses are similar to the
hierarchy and scale of down quark masses (see for example \cite{bs}),
one has also the relation $M_e \sim M_d$. Then a natural
assumption is $M_{\nu} \sim M_u$, in which case the
Dirac neutrino mass matrix can be written in the form
\ini
M_{\nu} \simeq \left( \begin{array}{ccc}
0 & a & 0 \\ a & b & c \\ 0 & c & 1
\end{array} \right) m_t,
\fin
where $a \ll b \sim c \ll 1$. The relation $b \simeq c$ in quark mass
matrices is discussed in Ref.\cite{fx}. We take $b$ and $c$ different but
of the same order. In fact, also matrices (3) and (5) can be written in the
form (6), with overall scales $m_b$ and $m_{\tau}$, respectively.

\section{Neutrino phenomenology}

Neutrino oscillation data imply that the lepton mixing matrix is given by
\ini
U \simeq \left( \begin{array}{ccc}
\frac{\sqrt2}{\sqrt3} & \frac{1}{\sqrt3} &
\epsilon \text{e}^{-\text{i} \delta} \\
-\frac{1}{\sqrt6} & \frac{1}{\sqrt3} & \frac{1}{\sqrt2} \\
\frac{1}{\sqrt6} & -\frac{1}{\sqrt3} & \frac{1}{\sqrt2}
\end{array} \right)
\text{diag}(\text{e}^{\text{i} \varphi_1/2},\text{e}^{\text{i} \varphi_2/2},1),
\fin
where $\epsilon < 0.16$, and the square mass differences among effective
neutrino masses $m_1,m_2,m_3$ are
\ini
\Delta m^2_{32} =m_3^2-m_2^2 \simeq 3 \cdot 10^{-3} \text{eV}^2,
\fin
\ini
\Delta m^2_{21} =m_2^2-m_1^2 \simeq 7 \cdot 10^{-5} \text{eV}^2.
\fin
In the basis where $M_e$ is diagonal, $M_L$ is obtained by the transformation
\ini
M_L = U^* D_L U^{\dg},
\fin
with $D_L=\text{diag}(m_1,m_2,m_3)$. The presence of phases $\varphi_1$,
$\varphi_2$ in the mixing matrix (7) is due to the Majorana nature of effective
neutrinos. In the lepton mixing matrix, $U_{\mu 3}$ is maximal, $U_{e2}$ is
large, and $U_{e3}$ is small. This is in contrast to the small quark mixings.

Since $\Delta m^2_{21} \ll \Delta m^2_{32}$, we may consider four kinds of
neutrino spectra: the normal hierarchy $m_1 \ll m_2 \ll m_3$, with
$m_3^2 \simeq \Delta m^2_{32}$ and $m_2^2 \simeq \Delta m^2_{21}$, the partial
degeneracy $m_1 \simeq m_2 \ll m_3$, with $m_3^2 \simeq \Delta m^2_{32}$,
the inverse hierarchy $m_1 \simeq m_2 \gg m_3$, with
$m_{1,2}^2 \simeq \Delta m^2_{32}$, and the almost degenerate spectrum
$m_1 \simeq m_2 \simeq m_3 \simeq 1$ eV. The elements of $M_L$ are given by
$$
M_{ee} \simeq \epsilon^2 m_3+\frac{m_2}{3}+2\frac{m_1}{3}
$$
$$
M_{e \mu} \simeq \epsilon \frac{m_3}{\sqrt2}+\frac{m_2}{3}-\frac{m_1}{3}
$$
$$
M_{e \tau} \simeq \epsilon \frac{m_3}{\sqrt2}-\frac{m_2}{3}+\frac{m_1}{3}
$$
$$
M_{\mu \tau} \simeq \frac{m_3}{2}-\frac{m_2}{3}-\frac{m_1}{6}
$$
$$
M_{\mu \mu} \simeq M_{\tau \tau} \simeq \frac{m_3}{2}+\frac{m_2}{3}+
\frac{m_1}{6}
$$
where phases are inserted by 
$\epsilon \ra \epsilon \text{e}^{\text{i} \delta}$,
$m_1 \ra m_1 \text{e}^{\text{i} \varphi_1}$,
$m_2 \ra m_2 \text{e}^{\text{i} \varphi_2}$,
and the relation $M_{\mu \mu} \simeq M_{\tau \tau}$ leads to the nearly maximal
mixing $U_{\mu 3}$.

Let us consider in particular the element $M_{ee}$, which is related to
neutrinoless double beta decay. For the normal hierarchy we obtain
(values in eV) $10^{-3} < M_{ee} \sim \sqrt{\Delta m^2_{21}} < 10^{-2}$,
for the partial degeneracy 
$10^{-3} < M_{ee} \sim 10^{-1} \sqrt{\Delta m^2_{32}} < 10^{-2}$,
for the inverse hierarchy
$10^{-2} < M_{ee} \sim \sqrt{\Delta m^2_{32}} < 10^{-1}$, and for the
degenerate spectrum $10^{-1} < M_{ee} < 1$. Hence, different spectra give
quite distinct prediction for $M_{ee}$.
There is a claim of evidence for the process \cite{kdhk},
with $M_{ee} =0.05-0.86$ eV,
in agreement with the degenerate spectrum and also the inverse hierarchy.
However, this result is controversial.

\section{The heavy neutrino mass matrix}

In this section we determine the right-handed neutrino mass matrix by means of
the inverse seesaw formula (2). We need $M_L^{-1}$, which is easily achieved,
since $M_L^{-1} =U D_L^{-1} U^T$.
We stress that the difference of $U_{e2}$ from the maximal mixing could be
ascribed to $M_e$ \cite{gt} and/or to renormalization \cite{run}.
Therefore, at the high scale and in the basis where $M_e$ is given by Eqn.(5),
we use the nearly bimaximal mixing in the seesaw,
\ini
U \simeq \left( \begin{array}{ccc}
\frac{1}{\sqrt2} & \frac{1}{\sqrt2} &
\epsilon \text{e}^{-\text{i} \delta} \\
-\frac{1}{2} & \frac{1}{2} & \frac{1}{\sqrt2} \\
\frac{1}{2} & -\frac{1}{2} & \frac{1}{\sqrt2}
\end{array} \right)
\text{diag}(\text{e}^{\text{i} \varphi_1/2},\text{e}^{\text{i} \varphi_2/2},1),
\fin
with $\epsilon \simeq 0$. Then the elements of $M_L^{-1}$ are given by
$$
M_{ee}^{-1} \simeq \frac{1}{2 m_1}+ \frac{1}{2 m_2}+ \frac{\epsilon^2}{m_3}
$$
$$
M_{e \mu}^{-1} \simeq -\frac{1}{2 {\sqrt 2} m_1}+
\frac{1}{2 {\sqrt2} m_2}+
\frac{1}{\sqrt2} \frac{\epsilon}{m_3}
$$
$$
M_{e \tau}^{-1} \simeq \frac{1}{2 {\sqrt 2} m_1}-
\frac{1}{2 {\sqrt2} m_2}+
\frac{1}{\sqrt2} \frac{\epsilon}{m_3}
$$
$$
M_{\mu \tau}^{-1} \simeq -\frac{1}{4 m_1}-\frac{1}{4 m_2}+\frac{1}{2 m_3}
$$
$$
M_{\mu \mu}^{-1} \simeq M_{\tau \tau}^{-1}
\simeq \frac{1}{4 m_1}+\frac{1}{4 m_2}+\frac{1}{2 m_3}
$$
where phases are inserted by $m_1 \ra m_1 \text{e}^{-\text{i} \varphi_1}$,
$m_2 \ra m_2 \text{e}^{-\text{i} \varphi_2}$,
$\epsilon \ra \epsilon \text{e}^{-\text{i} \delta}$.
Now, we determine the forms of $M_R$ according to the four kinds of mass spectra
for the effective neutrinos. We consider two extreme cases, that is
$\varphi_2 \simeq \varphi_1$ and $\varphi_2 \simeq \varphi_1+\pi$.
The other cases should be intermediate between those two.

For the normal hierarchy we obtain
\ini
M_R \simeq \left( \begin{array}{ccc}
a^2 & a(b-c) & -a \\ a(b-c) & (b-c)^2 & -(b-c) \\ -a & -(b-c) & 1
\end{array} \right) \frac{m_t^2}{4 m_1}.
\fin
An overall phase $\text{e}^{\text{i} \varphi_1}$ will be always absorbed.
The corresponding approximate form of $M_L$ at the low scale is given by
$$
M_L \sim \left( \begin{array}{ccc}
m_2 & m_2 & m_2 \\ m_2 & m_3 & m_3 \\ m_2 & m_3 & m_3
\end{array} \right).
$$

For the partial degeneracy, the case
$\varphi_2 \simeq \varphi_1$ leads to
$M_R$ the double of that in Eqn.(12). Instead,
$\varphi_2 \simeq \varphi_1 +\pi$ leads to the special form
\ini
M_R \simeq \left( \begin{array}{ccc}
0 & a & 0 \\ a & 2(c-b) & 1 \\ 0 & 1 & 0
\end{array} \right) \frac{a m_t^2}{2 {\sqrt2} m_1}.
\fin
The corresponding approximate forms of $M_L$ at the low scale are given by
$$
M_L \sim \left( \begin{array}{ccc}
m_{1,2} & m_2-m_1 & m_2-m_1 \\ m_2-m_1 & m_3 & m_3 \\ m_2-m_1 & m_3 & m_3
\end{array} \right),
$$
with $m_2-m_1 \sim \Delta m^2_{21}/m_{1,2}$, and
$$
M_L \sim \left( \begin{array}{ccc}
m_{1,2} & m_{1,2} & m_{1,2} \\ m_{1,2} & m_3 & m_3 \\ m_{1,2} & m_3 & m_3
\end{array} \right).
$$

For the inverse hierarchy
both cases $\varphi_2 \simeq \varphi_1$ and $\varphi_2 \simeq \varphi_1+\pi$
give
\ini
M_R \simeq \left( \begin{array}{ccc}
a^2 & a(b+c) & a \\ a(b+c) & (b+c)^2 & (b+c) \\ a & (b+c) & 1
\end{array} \right) \frac{m_t^2}{2 m_3}.
\fin
Note that while for the normal hierarchy the difference $(b-c)$ appears, for the
inverse hierarchy, instead, the sum $(b+c)$ appears. At the low scale we have
$$
M_L \sim \left( \begin{array}{ccc}
m_{1,2} & m_2-m_1 & m_2-m_1 \\ m_2-m_1 & m_{1,2} & m_{1,2} \\
m_2-m_1 & m_{1,2} & m_{1,2}
\end{array} \right),
$$
with $m_2-m_1 \sim \Delta m^2_{2,1}/m_{1,2}$, and a form of $M_L$ with
all entries of the order of $m_{1,2}$.

For the degenerate spectrum we get
in the case $\varphi_2 \simeq \varphi_1$
\ini
M_R \simeq \left( \begin{array}{ccc}
a^2 & ab & ac \\ ab & b^2+c^2 & c \\ ac & c & 1
\end{array} \right) \frac{m_t^2}{m_3}.
\fin
For $\varphi_2 \simeq \varphi_1 +\pi$ we have the same form as Eqn.(14).
At the low scale $M_L$ is of the same kind as the inverse hierarchy case. 

In the following sections we will consider, in a simplified approach,
the impact of $M_{\nu}$ and $M_R$ on the baryogenesis via leptogenesis
and the radiative lepton decays in some supersymmetric models. We first take
$M_{\nu} \sim M_u$, so that \cite{bs} 
\ini
M_{\nu} \sim \left( \begin{array}{ccc}
0 & \lambda^6 & 0 \\ \lambda^6 & \lambda^4 & \lambda^4 \\
0 & \lambda^4 & 1
\end{array} \right) m_t,
\fin
where $\lambda=0.22$ is the Cabibbo parameter. Since $b \sim c$,
we take only two forms for $M_R$,
one for the normal, inverse and degenerate case, and the other for the partial
degenerate case (13), that is 
\ini
M_R \sim \left( \begin{array}{ccc}
\lambda^{12} & \lambda^{10} & \lambda^6 \\
\lambda^{10} & \lambda^8 & \lambda^4 \\
\lambda^6 & \lambda^4 & 1
\end{array} \right) \frac{m_t^2}{m_k},
\fin
with eigenvalues $M_1/M_2 \sim \lambda^4$, $M_1/M_3 \sim \lambda^{12}$, and
\ini
M_R \sim \left( \begin{array}{ccc}
0 & \lambda^6 & 0 \\
\lambda^6 & \lambda^4 & 1 \\
0 & 1 & 0
\end{array} \right) \lambda^6 \frac{m_t^2}{m_1},
\fin
with eigenvalues $M_1/M_2 \sim \lambda^6$, $M_1/M_3 \sim \lambda^6$.
Notice that the scale of matrix (18) is smaller by several orders
with respect to the scale of matrix (17). Defining
$M_D=M_{\nu} U_R$, where $U_R$ diagonalizes $M_R$ ($M_D$ is the Dirac mass
matrix in the basis where $M_R$ is diagonal), we obtain $M_D^{\dg} M_D$,
which appears both in the formula for leptogenesis and in that for radiative
decays in supersymmetry,
\ini
M_D^{\dg} M_D \sim \left( \begin{array}{ccc}
\lambda^{12} & \lambda^{10} & \lambda^6 \\
\lambda^{10} & \lambda^8 & \lambda^4 \\
\lambda^6 & \lambda^4 & 1
\end{array} \right) {m_t^2},
\fin
\ini
M_D^{\dg} M_D \sim \left( \begin{array}{ccc}
\lambda^{12} & \lambda^{10} & \lambda^{10} \\
\lambda^{10} & 1 & 1 \\
\lambda^{10} & 1 & 1
\end{array} \right) {m_t^2}.
\fin
In the first case, matrix (17), we have $U_R$ near the identity and
$M_D^{\dg} M_D \sim M_R m_k$. In the other case, matrix (18),
$U_R$ is nearly unimaximal.
Therefore, in the matrix model made of (16) and (17), $M_{\nu}$ and $M_R$
give small mixings, so that large mixings in $M_L$ are produced through
a matching between $M_{\nu}$ and $M_R$ within the seesaw formula. Instead,
in the matrix model made of (16) and (18), the maximal mixing in $M_L$
comes from $M_R$. The structures (17) and (18) agree with the results of
Ref.\cite{ayu}, where it was realized that the seesaw enhancement of lepton
mixing can be achieved by strong mass hierarchy or large offdiagonal
elements in the heavy neutrino mass matrix.

\section{Baryogenesis via leptogenesis}

The baryogenesis via leptogenesis mechanism \cite{fy} is a well-known
mechanism for baryogenesis, related to the seesaw mechanism, where the decays
of heavy right-handed neutrinos produce a lepton asymmetry which is partly
transformed in a baryon asymmetry by electroweak sphaleron processes \cite{krs}.
The amount of baryon asymmetry is then given by the expression
\ini
Y_B \simeq \frac{1}{2} \frac{1}{g^*} ~d ~\epsilon_1,
\fin
where $\epsilon_1$ can be written as
\ini
\epsilon_1 \simeq \frac{3}{16 \pi}
\left[ \frac{(Y_D^{\dg} Y_D)_{12}^2}{(Y_D^{\dg} Y_D)_{11}} \frac{M_1}{M_2}+
\frac{(Y_D^{\dg} Y_D)_{13}^2}{(Y_D^{\dg} Y_D)_{11}} \frac{M_1}{M_3} \right],
\fin
see for instance Ref.\cite{ft}.
In these formulas $Y_D$ are Yukawa matrices, $g^* \simeq 100$, and $d < 1$ is a
dilution factor, which depends especially on the quantity
\ini
\tilde{m}_1 = \frac{(M_D^{\dg} M_D)_{11}}{M_1}.
\fin
Moderate dilution is present when $\tilde{m}_1$ is in the range of the
effective neutrino masses \cite{bp}.
The allowed value for the baryon asymmetry is $Y_B \simeq 9 \cdot 10^{-11}$,
see Ref.\cite{cfo}.
Yukawa matrices are obtained by dividing mass matrices by their overall scale.

For the two matrix models described in the previous section we get,
respectively,
\ini
\epsilon_1 \simeq \frac{3}{16 \pi}
\left(\frac{\lambda^{20}}{\lambda^{12}} \cdot \lambda^4 +
\frac{\lambda^{12}}{\lambda^{12}} \cdot \lambda^{12} \right)
\sim \frac{3}{16 \pi} \lambda^{12} \sim 10^{-10},
\fin
with $\tilde{m}_1 \sim m_k$, and
\ini
\epsilon_1 \simeq \frac{3}{16 \pi}
\left(\frac{\lambda^{20}}{\lambda^{12}} \cdot \lambda^6 +
\frac{\lambda^{20}}{\lambda^{12}} \cdot \lambda^{6} \right)
\sim \frac{3}{16 \pi} \lambda^{14} \sim 10^{-12},
\fin
with $\tilde{m}_1 \sim m_1$. Note that the two terms are comparable.
Moreover, it is clear that both models cannot provide a sufficient
amount of baryon asymmetry.

\section{Radiative lepton decays}

In supersymmetric seesaw models with universality above the heavy neutrino mass
scale, lepton flavor violations are produced by running effects from the
universality scale $M_U$ to the scale $M_R$ \cite{bm}.
The branching ratio for radiative lepton decays is given by the approximate
formula \cite{casas}
\ini
\text{Br}(l_i \ra l_j \gamma) \sim \frac{\alpha^3}{G_F^2 m_S^8}
\left(\frac{3 m_0^2+A_0^2}{8 \pi^2} \text{log} \frac{M_U}{M_R}\right)^2
(Y_D^{\dg} Y_D)_{ij}^2 ~\text{tan}^2 \beta,
\fin
with $l_1=e$, $l_2=\mu$, $l_3=\tau$. Here, $m_0$ is the universal scalar
mass, $A_0$ the universal trilinear coupling, and $m_S$ is the average slepton
mass at the weak scale, which can be quite different from $m_0$.
The experimental upper bounds are: 
$\text{Br}(\mu \ra e \gamma) < 1.2 \cdot 10^{-11}$,
$\text{Br}(\tau \ra e \gamma) < 2.7 \cdot 10^{-6}$,
$\text{Br}(\tau \ra \mu \gamma) <1.1 \cdot 10^{-6}$.
The first and third results are expected to be lowered by almost three orders
in the future.

Assuming $m_0=m_S=100$ GeV, $A_0=0$ and $\text{tan} \beta =50$, we obtain
for the first matrix model the values $10^{-18}$, $10^{-12}$, $10^{-9}$,
and for the second matrix model the values 
$10^{-18}$, $10^{-18}$, $10^{-3}$. 
Due to large uncertainties in supersymmetric parameters, we cannot make
definite predictions, so that previous numbers represent the effect of distinct
matrix models, which is our main interest here.
However, the element $(Y_D^{\dg} Y_D)_{32} \sim 1$ in matrix (20) seems
critical. 

\section{Discussion}

By inverting the seesaw formula we have calculated the heavy neutrino mass
matrix, and the implications for baryogenesis via leptogenesis and radiative
lepton decays in certain supersymmetric models. The analysis leads to two
distinct matrix forms, that is a nearly diagonal model
and a nearly offdiagonal model, which cannot provide sufficient
baryon asymmetry. For recent related studies, see Ref.\cite{other}.

We have assumed $M_{\nu} \sim M_u$. However, this assumption can be changed.
Indeed, the main feature of the Dirac neutrino mass matrix within the
seesaw mechanism is that its overall scale is of the order of $m_t$.
For example, we can take $M_{\nu} \simeq M_d m_t/m_b$, which means that it has
the same overall scale of $M_u$, but the internal hierarchy of $M_d$,
\ini
M_{\nu} \sim \left( \begin{array}{ccc}
0 & \lambda^3 & 0 \\ \lambda^3 & \lambda^2 & \lambda^2 \\
0 & \lambda^2 & 1
\end{array} \right) m_t.
\fin
In this case, sufficient baryon asymmetry is achieved, especially for
\ini
M_{R} \sim \left( \begin{array}{ccc}
\lambda^6 & \lambda^5 & \lambda^3 \\ \lambda^5 & \lambda^4 & \lambda^2 \\
\lambda^3 & \lambda^2 & 1
\end{array} \right) \frac{m_t^2}{m_k}.
\fin
The branching ratios of lepton decays are also enhanced
to $10^{-10}$, $10^{-7}$, $10^{-6}$. However, these strongly depend
on the mechanism of supersymmetry breaking.
In fact, in the previous section we have adopted a gravity mediated breaking,
where $M_U > M_R$, while for a gauge mediated breaking $M_U < M_R$ and running
effects are not induced.

An indication towards the existence of the seesaw mechanism would be the 
evidence for neutrinoless double beta decay. For the moment we predict
(in eV) $10^{-3} < M_{ee} < 0.86$. While the upper part of this range will
be checked rather soon, the lower part is more difficult to reach.

In conclusion, assuming baryogenesis from leptogenesis, we are led towards
a Dirac neutrino mass hierarchy similar to the down quark and charged lepton
mass hierarchy. In some supersymmetric scenarios, this model may be
checked by measurements of radiative lepton decays.

\end{document}